# The bending energy of a semi-flexible polymer chain and the polygons of the polymer chain

Pramod Kumar Mishra
*Department of Physics, DSB Campus, Kumaun University, Nainital (Uttarakhand)-263002, INDIA,*

**Abstract-** We consider random walk model of a semi-flexible polymer chain on a square and a cubic lattice to enumerate conformations of the polymer chain in two and three dimensions, respectively. The bending energy of the chain is assumed as the key factor which controls the minimum average length of the chain in between two successive bends in the chain; and the average length of the chain's per unit bend is defined as the persistence length of the polymer chain. It has been found that the minimum energy required to bend the chain is $\varepsilon_b = k_B * T * Log[2*(d-1)*g*l_p]$, where d, g and $l_p$ represents the dimensionality of the space, the step fugacity of the chain and the persistence length of the polymer chain, respectively.

I. INTRODUCTION

The long macromolecules play vital role in the living creatures and such molecules are also found in the non living items. In the case of living creature the *DNA* and the *Proteins* are the long macromolecules. The *DNA* and the *Proteins* are known for their good elastic properties. These Bio-molecules are soft objects and may be easily confined to fairly small space. The Bio-molecules are known to have variable range of stiffness. The macromolecules which require no energy to introduce bend in its segments are known as the flexible molecules, however, those molecules which require very large amount of the bending energy to produce even a single bend in its conformations are known as a stiff macro-molecule. Thus, the molecules which require intermediate value of bending energy to produce bend in its segments, and such molecules are known as the semi-flexible molecules. Such molecules are important and key to control Biological activities (replications, etc.) occurring in the living creatures [1-2].

The process of replication of the *DNA* or the *Proteins* is the key factor to Biological actions which are occurring in the living cells. The process of replication, gene regulations may take place easily provided the macromolecule has conformation like a loop (polygon). The looping of an amino acid chain is one of the key issues in the protein folding event. Thus the possibility of formation of the polymer loop (polygon) may be an interesting aspect which has been considered in the proposed study. The Biological macromolecules are heterogeneous and the distribution of different types of the monomers (repeat units) along the back bone of the chain is also random [1-4]. However, in our present model system we consider a linear homogenous polymer chain to describe the possibility of the formation of the polymer polygon on the basis of simple calculations.

The paper is organized as follows: In the section-2, we outline the polymer model of an ideal chain in two as well as three dimensions. Since, we consider polymer polygon which has close loop structure and therefore, we consider the chain as an ideal polymer chain. The method of recursion relations were discussed in brief to calculate the partition function of the polymer chain in two as well as three dimensions using square and cubic lattices, respectively. In the section three, we calculate the persistence length of the polymer chain and derive expression for the bending energy of the chain conformations which are forming close loop (polygons). We conclude the discussion in the section four by summarizing our findings.

II. MODEL AND METHODS

The random walk is used on a square and a cubic lattice to mimic the polymer chain conformations. The walker begins from a point and takes steps along all possible directions to mimic the polymer chain conformations and only those conformations were taken into considerations which have end monomer on the site where the polymer chain is grafted. In this case, since the close loops (polygons, as schematically shown in Figure No. 1) of the chain were considered therefore the chain is modeled as an ideal polymer. The walker, thus can take steps along $\pm x$ and $\pm y$ directions in the case of square lattice (i. e. in the case of two dimensional model) while in the case of cubic lattice the walker may take steps along $\pm x$, $\pm y$ and $\pm z$ directions, when the chain conformations are considered in three dimensions [5].

The partition function of the polymer chain may be written as [5-11],

$$Z(g,k) = \sum_{N=0}^{\infty} \sum_{N_B=0}^{\infty} g^N k^{N_B} \quad \ldots\ldots\ldots\ldots\ldots(1)$$

The persistence length ($l_p$) of the polymer chain may be defined as the average length of the polymer chain per unit bend and it may be calculated from the following relations:

$L_p = <L>/<N_B>$

$$= [\frac{\partial Log(Z(g,k))}{\partial Log(g)}] \Big/ [\frac{\partial Log(Z(g,k))}{\partial Log(k)}] \quad \ldots\ldots\ldots(2)$$

It is also to be noted here that we consider those conformations of the polymer chain which has starting and end monomers at the site where the polymer chain is grafted. The perimeter of the polymer polygon is 4n monomers (where n=$l_p$) i. e. the n monomers are along each of the direction so that the polymer polygon may be closed one, where n=1, 2, 3…., $l_p$ monomers.

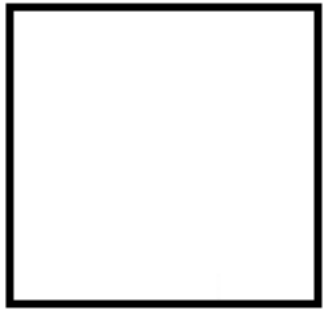

Figure No. 1: We have shown the polymer polygon schematically. Each arm of the polygon of the polymer has $l_p$ monomers. Thus the perimeter of the polygon is 4*$l_p$ while area of the polymer polygon is $l_p$*$l_p$.

## III. CALCULATIONS

We calculate the average length of the polymer chain and also the average number of the bends in the chain so that we may have value of the persistence length of the polymer chain using equation nos. (1) and (2). It has been found from these calculations that the bending energy of the chain conformations in two dimensions may vary with polygon size ($n=l_p$) as,

$$\mathcal{E}_b = k_B * T * \text{Log}[2*g*l_p] \quad\quad\quad (3)$$

While, in the case of three dimensions or for the case of a cubic lattice the bending energy of the polymer chain may be expressed as (as shown in Figure No. 2),

$$\mathcal{E}_b = k_B * T * \text{Log}[4*g*l_p] \quad\quad\quad (4)$$

## V. RESULTS AND THE CONCLUSIONS

The random walk lattice model is used to mimic conformations of a long linear semi-flexible polymer chain in two and three dimensions using square and cubic lattices, respectively. The polymer chain is grafted on a point and all those conformations of the polymer chain were considered which have starting as well as end monomers on the site where the polymer chain is grafted. It has been found that there are 8 polygons (each polygon has the perimeter $4*l_p$ and the area is $l_p^2$) in the case of two dimensions and while there are 12 polygons (each polygon has the perimeter $4*l_p$ and the area is $l_p^2$) which are found to have first as well as the last monomer on the site where the polymer chain is grafted. The probability of finding such polymer polygon in two dimensions is $8/(4*l_p)^4$ and in three dimensions the probability is $12/(4*l_p)^6$. The value of $l_p$ is 1, 2, 3, ......, n monomers.

The standard methods of the statistical Physics were used to calculate the persistence length of the polymer chain. It has been found in the case of two as well as three dimensions that the bending energy of the chain may be expressed as the polygon size ($l_p$) as,

$$\mathcal{E}_b = k_B * T * \text{Log}[2*(d-1)*g*l_p]$$

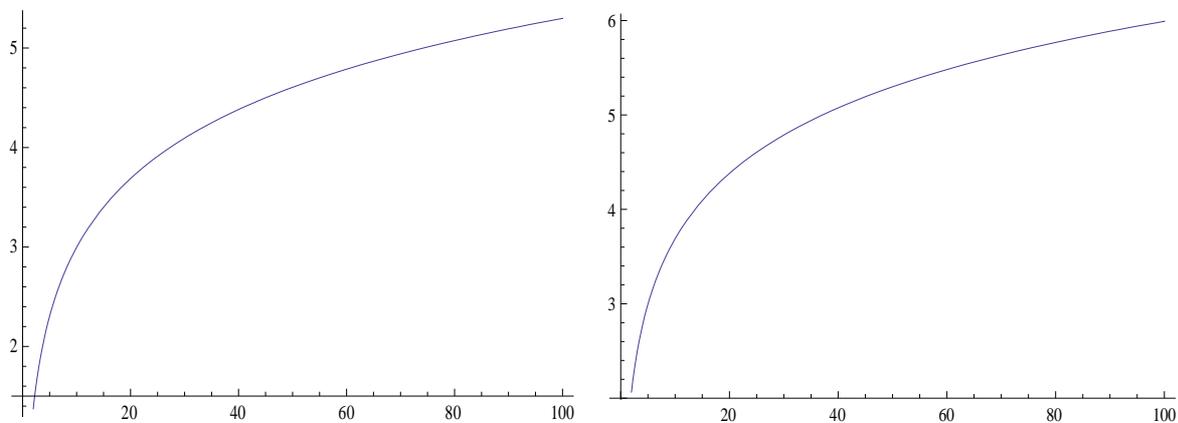

Figure No. 2: The variation of ($\mathcal{E}_b/k_B*T$) versus Log [$2*g*l_p$] is shown for two dimensional case, while the variation of ($\mathcal{E}_b/k_B*T$) versus Log[$4*g*l_p$] is also shown for the case of three dimensions in this figure (here g=1 unit).